\documentclass{desyproc}
\usepackage{color}
\definecolor{purple}{rgb}{0.4,0,0.6}

\begin{document}
\title{\vspace{-3cm}{\small \hfill{DESY 10-139; IPPP/10/70; DCPT/10/140}}\\[1.8cm]
Improving the Discovery Potential of Future\\ Light-Shining-through-a-Wall Experiments}

\author{{\slshape Paola Arias$^1$, Joerg Jaeckel$^2$, Javier Redondo$^3$, Andreas Ringwald$^1$}\\[1ex]
$^1$Deutsches Elektronen-Synchrotron, Notkestra\ss e 85, D-22607 Hamburg, Germany\\
$^2$Institute for Particle Physics Phenomenology, Durham University, Durham DH1 3LE, UK\\
$^3$Max Planck-Institut f\"ur Physik, F\"ohringer Ring 6, D-80805 M\"unchen, Germany}

\contribID{Arias\_Paola}

\desyproc{DESY-PROC-2010-03}
\acronym{Patras 2010} 
\doi  

\maketitle

\begin{abstract}
Planning for the next generation of light-shining-through-wall experiments has started. It is therefore timely to investigate possible ways to optimize their setups. The goals are to improve the sensitivity towards smaller couplings and increase the mass range to which the experiments are sensitive. We discuss possible magnet arrangements and the effects of the unavoidable gaps in the magnetic field profile. Furthermore, we discuss requirements on the diameter of the laser beam and aperture of the magnets in order to achieve high-quality cavities.

\end{abstract}

\noindent
Several well motivated extensions of the Standard Model predict the existence of very weakly interacting slim particles (dubbed WISPs). In particular, top-down models arising from compactifications in string theory suggest plenty of
them~\cite{Witten:1984dg
}, such as  axions~\cite{Peccei:1977hh
},
axion-like particles (ALPs), hidden U(1) gauge bosons~\cite{Okun:1982xi} and mini-charged particles~\cite{Holdom:1985ag}. Optical precision experiments are a powerful tool to search for these particles. In particular, a class of very simple and effective laser experiments is based on photon -- WISP -- photon oscillations: the so called {\em light-shining-through-a-wall} (LSW) experiments.\\
The sensitivity of LSW experiments has grown considerably over the last few years. In this work we will concentrate on ALPs, whose  coupling to two photons has been recently constrained, using the LSW technique, to be smaller than $g\sim \mbox{few} \times 10^{-7}\,{\rm GeV}^{-1}$~\cite{Ehret:2010mh}.
The goal of the next generation of LSW experiments should be to surpass the present
limits of $g\sim \mbox{few} \times 10^{-10}\,{\rm GeV}^{-1}$ obtained by CAST, the CERN Solar Axion Telescope~\cite{cast}.
The most straightforward way to increase the sensitivity for ALPs is to enlarge the product $B L$ of the magnetic field strength ($B$) and the length of the magnetic region ($L$), since the probability of $\gamma\to {\rm ALP}\to \gamma$ conversion,
at small masses, scales as $\propto (BL)^4$. Nearly all of the current generation of LSW experiments (ALPS~\cite{Ehret:2010mh,alpscollaboration, Ehret:2009sq}, GammeV~\cite{gammev}, LIPSS~\cite{lipss} and OSQAR~\cite{osqar}), recycle one or two of the long superconducting dipole
magnets from accelerator rings, such as the ones from HERA, Tevatron or LHC. Additional improvements can be
achieved by: using a larger number of these magnets~\cite{Ringwald:2003nsa}, progress in laser and detector technology, and the
introduction of matched optical resonators in both, production and regeneration
regions~\cite{Hoogeveen:1990vq
}.
Implementing these advances, sensitivities in the $g\sim {\rm few}\times 10^{-12}$~GeV$^{-1}$ range seem achievable for light ALPs, thus opening great opportunities for discoveries.

However, using an array of magnets to increase the magnetic field region modifies the production/regeneration form factors
and high quality cavities must have a minimum diameter in order to avoid excessive clipping losses, in turn requiring magnets with sufficiently large aperture.
We will discuss these effects and ways to profit from them or at least minimize their negative impact.
Finally, we want to find an optimal configuration, based on existing technology, that maximizes the sensitivity of the experiment for a wide range of ALPs masses.

\section{Oscillation probability for realistic magnet arrangements}

To start with, let us recall that the probability of a laser photon of frequency $\omega$ converting into an ALP of mass $m_\phi$ - and vice versa - after traveling a distance $L$, is given by
\[
 P_{\gamma\rightarrow\phi}=P_{\phi\rightarrow\gamma}=\frac{1}4\frac{\omega}{k_\phi}\left(gBL\right)^2
|F(qL)|^2,
\label{prob}
\]
where $q\equiv|k_\gamma-k_\phi|$ is the difference of the photon and ALP momenta which for small $m_\phi$ is approximately $q\approx m_\phi^2/(2\omega)$.
The function $F$, sometimes called form-factor, characterizes the profile of the
magnetic field, which we wrote as $\vec B(\vec x)= \vec e_z B f(x)$, along the photon trajectory,
\[
F(qL)\equiv \frac{1}{L}\int_0^L dx' f(x')~ e^{iqx'} .
\]
In the current generation of LSW experiments,
exploiting single dipole magnets with homogeneous $B$, both on the generation and regeneration side,
the form factor takes the  familiar form
$
F_{\rm single}(qL)= (2/(qL)) \sin\left( qL/2\right).
$
The maximum conversion probability
$P_{\gamma\to\phi } =
g^2B^2L^2/4$
is achieved in vacuum for small momentum transfer,  $qL/2\ll 1$, corresponding to small masses, where the form factor takes its maximum value, $F_{\rm single}(qL)=1$. Thus, the goal is to optimize the setup such that $F(qL)$ is as close to 1 as possible, for a wide range of $m_\phi\ll\omega$.\\
The setup foreseen for the next generation of LSW experiments will exploit series of $N$ dipole magnets (see e.g. \cite{alpscollaboration, gammev}), including a natural and probably unavoidable ``gap'', with no magnetic field in between each magnet. Therefore, we should re-write the form factor for a longitudinal profile corresponding to $N$ equally spaced magnets, each of length $\ell$, separated from each other by a fixed length $\Delta$. In fact, a short calculation results in
\begin{equation}
F_{N,\Delta}(qL)
=\frac{1-e^{iqN(\ell+\Delta)}}{1-e^{iq(\ell+\Delta)}} \times \frac{F_{\rm single}(q \ell)}{N}
=\frac{2}{qL}\sin\left(\frac{qL}{2N}\right)\frac{\sin\left(\frac{qN}{2}\left(L/N+\Delta\right)\right)}
{\sin\left(\frac{q}{2}\left(L/N+\Delta\right)\right)},
\label{alternating1}
\end{equation}
with $L=N\ell$ the total length of the magnetic field.
The effect of the gap in between the magnets acts as a phase
added to a single magnet just as the \emph{phase shift plates} proposed in~\cite{Jaeckel:2007gk}, but with the advantage that the gaps  introduce no optical losses. Besides the usual zeros of $F_{\rm single}(qL)$, located at
$m_\phi^2=4k \pi \omega/\ell$, with $k \in \mathbb Z^+$, other zeros (gap dependent) appear in $F_{N,\Delta}(qL)$, given by $m_\phi^2=4k\pi  \omega/\left(N\left(\ell+\Delta\right)\right)$. Therefore, in this setup, the loss of the coherent photon-ALP conversion occurs already at somewhat smaller masses.

The latter issue can be ameliorated by considering an array on $N$ identical magnets of length $\ell$, segmented into $n$ subgroups of alternating polarity~\cite{VanBibber:1987rq}, such that the total magnetic length is given by $L=N\ell$.
Including a fixed gap in-between the magnets, as we did before, we find that again the form factor is modified with an extra oscillation mode,
\begin{eqnarray}
F_{N,n,\Delta}(qL)=\left\{ \begin{array} {l l}
\frac{2}{qL} \sin\left(\frac{qL}{2N}\right) \frac{\sin\left(\frac{qN}2(L/N+\Delta)\right)}{\sin\left(\frac{q}{2}(L/N+\Delta)\right)}
\tan\left(\frac{qN}{2n}(\frac{L}N+\Delta)\right), \ \ \ \mbox{$n$ even},\\
\frac{2}{qL} \sin\left(\frac{qL}{2N}\right) \frac{\cos\left(\frac{qN}2(L/N+\Delta)\right)}{\sin\left(\frac{q}{2}(L/N+\Delta)\right)}
\tan\left(\frac{qN}{2n}(\frac{L}N+\Delta)\right), \ \ \ \mbox{$n$ odd}.\\
\end{array}\right.
\label{alternating2}
\end{eqnarray}
In the limit $\Delta \rightarrow 0$, the formulas from~\cite{VanBibber:1987rq} are recovered.\\
We can now maximize Eqs.~(\ref{alternating1}) or (\ref{alternating2}) varying the gap length. We are then able -- in principle -- to optimize the sensitivity for given values of $m_\phi$ changing the size of the gap, scanning optimally the ALP parameter space. For instance, maximization of Eq.~(\ref{alternating1}) gives $(q\ell/2)\left(1+\Delta_{\rm opt}/\ell\right)=k\pi$, with $k \in \mathbb Z$. Unfortunately, using this equation, a full scan of the parameter region it is not possible experimentally, because  $\Delta$  is limited by the length of the setup, and in particular also by the maximal length of the cavity (see below).\footnote{As can be seen from the maximization condition, for small $q\ell$ the size of the gap grows.}

\section{Discovery potential}

\begin{wraptable}{r}{0.31\textwidth}
\vspace{-0.4cm}
\scriptsize{\centerline{\begin{tabular}
{|l||l|}
    \hline
$\ell$ & $14.3~\mbox{m} \ \ (6\times 6)$\\
\hline
$B$ & $9.5$~T \\
\hline
${\mathcal P}_{\rm prim}$ & 30~W \\
\hline
$\mathcal F_g=\mathcal F_r$ &  $\pi \times 10^{5}$\\
\hline
$\omega$ & 1.17~eV\\
\hline
$\tau$ & 50~h\\
\hline
$\eta $ & 0.95 \\
\hline
\end{tabular}}}
\vspace{-0.1cm}
\caption{\scriptsize{Benchmark values for a next generation of LSW experiment.}}
\vspace{-0.4cm}
\label{table:table1}
\end{wraptable}
To estimate the sensitivity of next generation LSW experiments,  we have chosen a benchmark set of reasonably realistic input parameters (summarized in Table~\ref{table:table1}). Here, $\mathcal{P}_{\rm prim}$ is the primary laser power, ${\mathcal F}_{g,r}$ are the finesses of the generation and regeneration cavities, $\tau$
is the measurement time,
and $\eta$ is the spatial overlap integral between the ALP mode and the electric field mode.

For instance, let us assume a $6+6$ LHC setup (i.e. 6 magnets in the production and in regeneration side, respectively) and the benchmark parameters given by Table~1. Using the estimated number of photons in the regeneration cavity, given in Ref.~\cite{Mueller:2009wt}, such configuration can reach a limit of $g \sim 5 \times 10^{-12} $~GeV$^{-1}$ for small masses.\footnote{
We have assumed negligible dark count rate in the detector, therefore the sensitivity to $g$ grows as $\tau^{-1/4}$.}
Figure~1 a) shows the impact of considering a realistic configuration with gaps in-between the magnets versus neglecting them. As we can see, enlarging the magnetic region of LSW experiment by placing a series of superconducting magnets modifies the probability of  photon-ALP-photon conversion. Nonetheless, the unavoidable gaps in-between the magnets can represent an improvement in the setup, since they appear as a {\em free} parameter, adding an extra oscillation mode.

However, care must be taken: the enlargement of the resonant cavity is strongly dependent on the diameter of the laser beam and therefore, the aperture of the magnet.
A rough estimate of the optimum cavity length, as a function of the aperture of the magnet~($a$) and the wavelength of the laser~($\lambda$), gives  $L_{\rm opt}=(0.42)^2\, (\pi a^2)/(2 \lambda)$~\cite{paper}. For instance, using LHC magnets ($a\approx$ 28~mm) the optimum cavity length is $L_{\rm opt} \sim 204$~m.

Finally, in Fig.~1 b) we display the projected sensitivity on ALPs for the next generation of LSW experiments. With six magnets on each side of the experiment we can produce four different symmetric configurations: $\uparrow
\uparrow\uparrow\uparrow \uparrow\uparrow$, $\uparrow \downarrow\uparrow \downarrow\uparrow \downarrow$,
$\uparrow\uparrow \downarrow\downarrow\uparrow\uparrow$, $\uparrow\uparrow\uparrow\downarrow\downarrow\downarrow$. Moreover, we have calculated the sensitivity for seven different gap sizes. We infer that, using these alternating field configurations, we are able to restore the sensitivity almost completely
up to masses of order $m\sim 5\times 10^{-4}$~eV.

\begin{figure}
\vspace{-0.5cm}
   \includegraphics[width=0.5\textwidth]{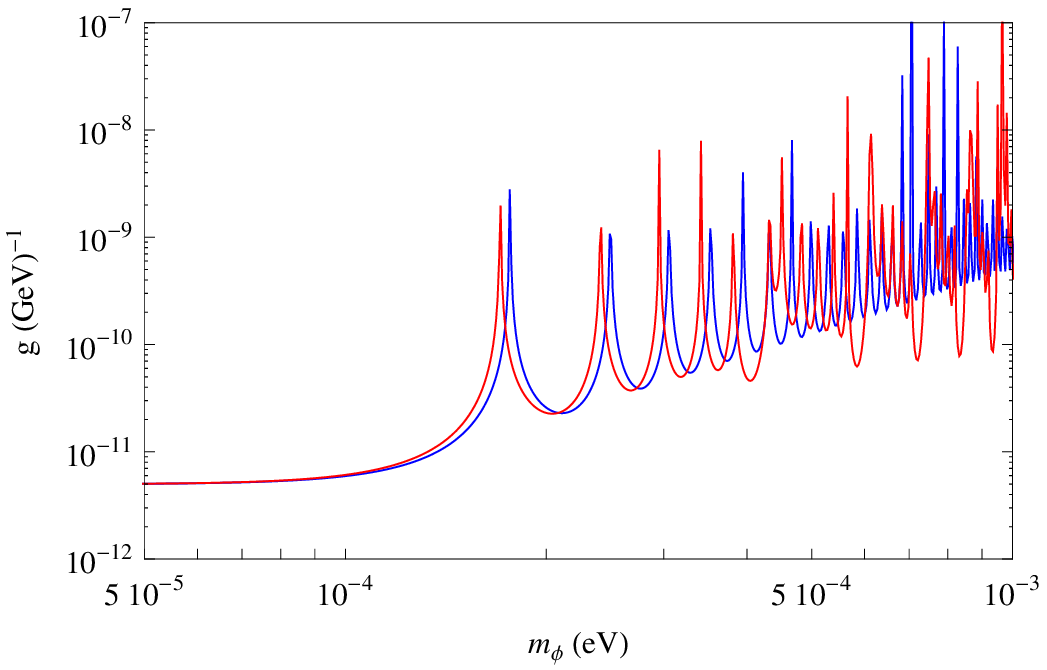}
\vspace{-0.1cm}
   \includegraphics[width=0.5\textwidth]{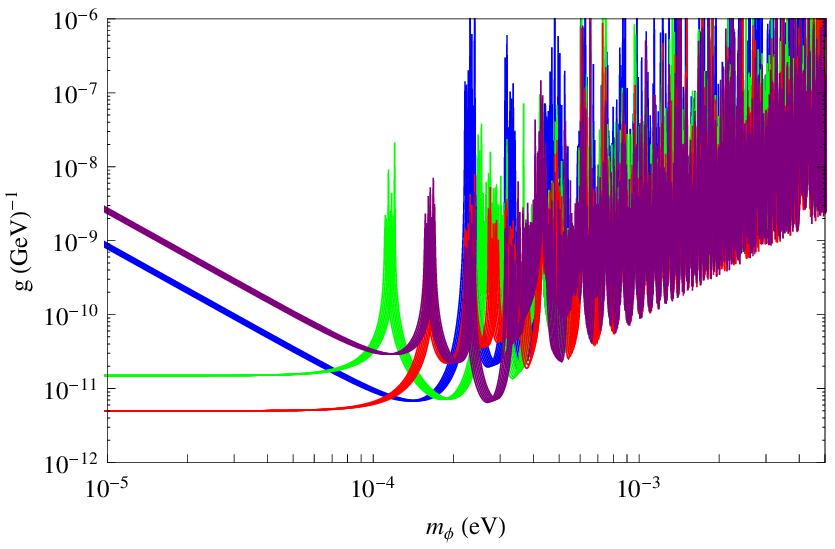}
\Text(-170,122)[l]{ \footnotesize{\color{red}{$\uparrow\uparrow\uparrow\uparrow\uparrow\uparrow$}}}
\Text(-170,110)[l]{ \footnotesize{\color{blue}{$\uparrow\uparrow\uparrow\downarrow\downarrow\downarrow$}}}
\Text(-170,98)[l]{ \footnotesize{\color{green}{$\uparrow\uparrow\downarrow\downarrow\uparrow\uparrow$}}}
\Text(-170,86)[l]{ \footnotesize{\color{purple}{$\uparrow\downarrow\uparrow\downarrow\uparrow\downarrow$}}}
\vspace{-0.2cm}
  \caption{\scriptsize{a) Blue line represents the sensitivity of a setup without taking into account the gap in-between the magnets $\left(F(qL)=F_{\rm single}(qL)\right)$. Red line corresponds to the same setup, considering $\Delta=1$~m. b) The four configurations complement each other, fullfilling the sensitivity for small masses. With $\Delta [m]=\left\{ 3.83, 3.30, 2.80, 2.33, 1.89, 1.47, 1.08\right\}.$}}
\vspace{-0.2cm}
  \label{fig:constraints}
\end{figure}

\section{Conclusions}
We have shown the modifications on the form factor, when a series of magnets with gaps in-between is exploited in LSW experiments. With a setup of $6+6$ LHC magnets, a sensitivity of $g\sim 5\times 10^{-12}$~GeV$^{-1}$ seems realistic, for $m_\phi \leq 5 \times 10^{-4}$~eV. The next generation of LSW experiments may therefore indeed explore territory in parameter space that has not been excluded yet by astrophysics and cosmology. Moreover, it may probe ALP interpretations of hints for cosmic photon regeneration~\cite{roncadelli} and for an anomalous energy loss in white dwarfs~\cite{Isern}.


\begin{footnotesize}

\end{footnotesize}


\end{document}